\documentclass[prd,aps,showpacs,nofootinbib,superscriptaddress,floatfix]{revtex4}
\usepackage{epsfig}
\usepackage{amsmath, slashed}
\usepackage{color}

\begin{document}

\title{ Singularities in Twist-3 Quark Distributions}

\author{Fatma P. Aslan and Matthias Burkardt}
 \affiliation{Department of Physics, New Mexico State University,
Las Cruces, NM 88003-0001, U.S.A.}
\date{\today}

\begin{abstract}
 We find in one-loop calculations and spectator models that twist-3 GPDs exhibit discontinuities. In the forward limit  these discontinuities grow into Dirac delta functions which are essential to satisfy the sum rules involving twist-3 PDFs. We calculate twist-3 quasi-PDFs as a function of longitudinal momentum and identify the Dirac delta function terms with momentum components in the nucleon state that do not scale as the nucleon is boosted to the infinite momentum frame.
\end{abstract}

\maketitle

\section{Introduction} 
\label{s:intro}
A  complementary picture of the  nucleon structure is obtained by simultaneous  information on both transverse spatial and longitudinal momentum distributions of partons. The relevant physical observables are generalized parton distributions (GPDs) ~\cite{ Mueller:1998fv,Radyushkin:1996nd, Ji:1996nm, Radyushkin:1996ru}. Theoretically they are calculated from nonforward  matrix elements of local operators, and experimentally they are accessible through exclusive deep inelastic scattering  experiments, such as deeply virtual Compton scattering (DVCS) ~\cite{ Ji:1996nm, Collins:1998be}. GPDs give information about the spin, momentum and spatial distribution  of the quarks, anti-quarks and gluons within a fast moving nucleon \cite{Burkardt:2000za,Diehl:2002he,Burkardt:2002hr} and therefore provide a remarkable insight on its inner structure.

One property to classify the GPDs, is their twist \cite{Jaffe:1996zw}. Twist determines the order in $Q^2$ (squared four-momentum transfer) at which a matrix element contributes to the physical amplitude of a given hard process. With increasing twist the number of partons which participate in that matrix element also tend to increase. At leading twist, twist-2 GPDs describe two-particle correlations in the nucleon while the next leading twist, twist-3 GPDs also involve three-particle correlations such as quark-gluon-quark (qgq). It is advantageous to define the twist in the infinite momentum frame (IMF) where the nucleon has a large momentum in the longitudinal direction (direction of the nucleon propagation), i.e. $P^+\gg$ M and nearly zero momentum in the transverse direction, i.e. $P_T\approx 0$ \cite{Weinberg:1966jm}. In the IMF the twist of a distribution can be identified with its  behavior under a longitudinal momentum boost. While twist-2 distributions are invariant under the boosts along the longitudinal direction, twist-3 distributions scale as  $1/P^+$.

 In the Bjorken limit, the  matrix elements are dominated by twist-2 operators \cite{Bjorken:1968dy}. Even though they are mostly relevant for subleading corrections, there are several motivations to study twist-3 GPDs. 

\begin{itemize}
  \item They involve a novel type of information on qgq correlations which is not contained in  twist-2 distributions. These qgq correlations can be interpreted as an average transverse force acting on the quarks inside a nucleon. The new information embodied in twist-3 GPDs is the distribution of that force on the transverse plane \cite{Burkardt:2008ps}.
 
    \item Twist-3 corrections may not be negligible in the upcoming detailed measurements of DVCS amplitude at 12 GeV in Jefferson Lab.
   
    \item  Twist-3 GPDs may provide an alternative source of information on orbital angular momentum of the quarks through the sum rule which relates the second moment of a particular twist-3 GPD, $G_2$ and kinetic angular momentum of the quarks inside a longitudinally polarized nucleon \cite{Penttinen:2000dg, Ji:1996ek},
\begin{align}  \label{eq:polyakovsumrule}
L_{kin}^q=-\int{} dx  x  G_2 ^q  (x,\xi=0,t=0),
\end{align}
where $x$ is the average longitudinal quark momentum, $\xi$ is the longitudinal and $t$ is the total momentum transfer to the nucleon.
 
\end{itemize}

Twist-3 GPDs generically exhibit discontinuities at the points of particular interest and importance ($x =\pm\xi$). These points correspond
to configurations in which one of the partons has a vanishing momentum component in the matrix element describing the scattering amplitude.  There are several studies \cite{Belitsky:2000vx,Kivel:2000rb,Radyushkin:2000jy, Kivel:2000cn, Radyushkin:2000ap, Belitsky:2000vk, Kivel:2000fg, Anikin:2001ge, Belitsky:2001yp, Kivel:2001rw, Belitsky:2001ns, Kiptily:2002nx, Kivel:2003jt, Freund:2003qs} which reveal the discontinuties of twist-3 GPDs using Wandzura-Wilczek (WW) approximation \cite{Wandzura:1977qf}. However, we show that without using WW approximation twist-3 GPDs are also discontinuous in quark target model (QTM) and scalar diquark model (SDM). Potentially such discontinuities lead to divergent scattering amplitudes and endanger the factorization  of the hard-scattering
process \cite{Collins:1989gx}. However in Ref.\cite{Aslan:2018zzk}   it has been shown that the discontinuities cancel for the linear combinations of twist-3 GPDs which enter the DVCS amplitude  and therefore twist-3 amplitudes are
consistent with DVCS factorization.
Even though they do not exhibit a problem for factorization we will show that  in the forward limit these discontinuties can grow into Dirac delta functions which are essential for satisfying relevant Lorentz invariance relations.  Investigating twist-3 quasi-PDFs as a function of longitudinal momentum reveals that the Dirac delta function terms correspond to momentum components in the nucleon state that do not scale  as the nucleon is boosted to the IMF.

There are several parametrizations  of the correlators which define GPDs \cite{Belitsky:2001yp,Belitsky:2001ns,Meissner:2009ww}. The relations between the different parametrizations are given in Ref.\cite{Aslan:2018zzk}.  We use the following parametrization \cite{Kiptily:2002nx}   and adopt the light front gauge $A^+=0$ where the Wilson lines can be ignored \cite{Ivanov:1985np},

\begin{eqnarray}
\label{eq:vector}
\dfrac{1}{2}\int{}\dfrac{dz^-}{2 \pi} e^{i x p^+z^-}\langle P',S'|\:\overline{q}(-\dfrac{z^-}{2}) \gamma^j q (\dfrac{z^-}{2})\:|P,S\rangle \\ \nonumber
=\dfrac{1}{2p^+}\overline{u}(P',S')\Big[\dfrac{\Delta_{\perp}^j}{2M}G_1+\gamma^j (H+E+G_2)
+\dfrac{\Delta_{\perp}^j}{p^+}\gamma^+ G_3+\dfrac{i \epsilon_T^{jk}\Delta_{\perp}^k}{p^+}\gamma^+\gamma_5G_4\Big]{u}(P,S),
\end{eqnarray}

\begin{eqnarray}
\label{eq:axialvector}
\dfrac{1}{2}\int{}\dfrac{dz^-}{2 \pi} e^{i x p^+z^-}\langle P', S'|\:\overline{q}(-\dfrac{z^-}{2}) \gamma^j \gamma_5q (\dfrac{z^-}{2})\:|P,S\rangle \\ \nonumber
=\dfrac{1}{2p^+}\overline{u}(P',S')\Big[\dfrac{\Delta_{\perp}^j}{2M}\gamma_5(\widetilde{E}+\widetilde{G}_1)+\gamma^j\gamma_5 (\widetilde{H}+\widetilde{G}_2) 
+\dfrac{\Delta_{\perp}^j}{p^+} \gamma^+\gamma_5\widetilde{G}_3+\dfrac{i \epsilon_T^{jk}\Delta_{\perp}^k}{p^+}\gamma^+\widetilde{G}_4\Big]{u}(P,S).
\end{eqnarray}

In Eqs.(\ref{eq:vector}) and (\ref{eq:axialvector})  $P (P')$ is the incoming (outgoing), $p^+$ is the average longitudinal nucleon four-momentum, $S (S')$ is the initial (final)  nucleon spin, $M$ is the nucleon mass and $H, E, \widetilde{H}, \widetilde{E}$ are twist-2,  $G_1,...,G_4, \widetilde{G}_1,...,\widetilde{G}_4$ are twist-3 GPDs.

In this study, we focus on the twist-3 GPDs, $G_2$ and  $\widetilde{G}_2$ since they are essential in the sense that $G_2$  is related to the quark kinetic orbital angular momentum  via Eq.(\ref{eq:polyakovsumrule})  and  $\widetilde{G}_2$ reduces to $g_2(x)$ in the forward limit which enters the polarized DIS cross section \cite{Jaffe:1990qh}. 
This paper is organized as follows:
In section \ref{section:QTM} and \ref{section:SDM}, the twist-3 GPDs,  $G_2$  and $\widetilde{G}_2$ are calculated in QTM and SDM respectively. The discontinuities of  $G_2$  and $\widetilde{G}_2$ and their behavior under decreasing longitudinal momentum transfer are investigated. In section \ref{section:pdf} twist-3 PDF, $g_2(x)$ and quasi-PDF, $g_2^{quasi}(x)$ are calculated using SDM. The discontinuities of $\widetilde{G}_2$ in the forward limit are identified with Dirac delta function terms in $g_2(x)$. Calculation of $g_2^{quasi}(k^z)$  reveals that these Dirac delta function terms correspond to the momentum components in the nucleon state that do not scale as the nucleon is boosted to the IMF. As shown in section \ref{section:sumrules} neclecting  the Dirac delta functions would lead to an apparent violation of sum rules for twist-3 PDFs and GPDs.  In section \ref{section:sumary} our work is summarized.

\section{ $G_2$ and $\widetilde{G}_2$ in Quark Target Model}\label{section:QTM}
\begin{figure}[ht]
    \centering
    \includegraphics[width=8.cm]{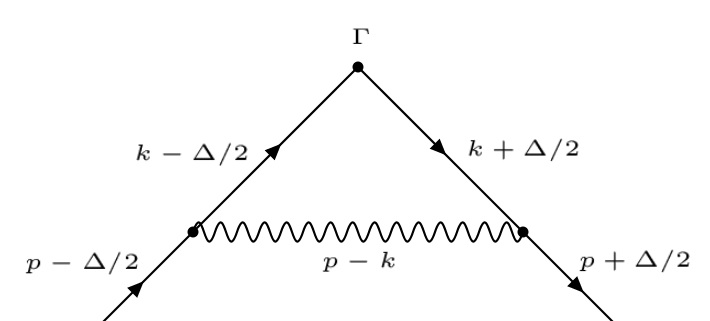}
    \caption{Quark target model in a symmetric frame.}
    \label{fig:QTM}
\end{figure}

In QTM a quark or an anti-quark releases a photon/gluon and recombines with it after the interaction. FIG.\ref{fig:QTM} shows QTM in a symmetric frame where the kinematic variables are,

$\Delta$: The four-momentum transfer, 

$ P=p-\dfrac{\Delta}{2}  (P'=p+\dfrac{\Delta}{2})$: The incoming (outgoing) four-momentum,

$  p$: The average momentum (with $ p_{\perp}=0$),

$k-\dfrac{\Delta}{2} (k+\dfrac{\Delta}{2})$: The four-momentum before (after) the interaction. 

To calculate $G_2$,  the matrix element on the left-hand side of Eq.(\ref{eq:vector})  is written using QTM with the vertex operator $\Gamma=\gamma^{\perp},$

\begin{eqnarray}\label{vectorQTM}
-\dfrac{ig^2}{2}\int{}\dfrac{d^4k}{(2 \pi)^4} \delta(k^+-xp^+)\overline{u}(P',S')\gamma^{\mu}\dfrac{(k\!\!\!/+\dfrac{\Delta\!\!\!/}{2}+m)}{[(k+\dfrac{\Delta}{2})^2-m^2+i \epsilon]} \gamma^{\perp} 
  \dfrac{(k\!\!\!/-\dfrac{\Delta\!\!\!/}{2}+m)}{[(k-\dfrac{\Delta}{2})^2-m^2+i\epsilon]}\gamma^{\nu}\\ \nonumber
 \times \Big[g_{\mu\nu}  -\dfrac{n_{\nu}(p_{\mu}-k_{\mu})}{p^+-k^+}-\dfrac{n_{\mu}(p_{\nu}-k_{\nu})}{p^+-k^+}\Big] \dfrac{1}{[(p-k)^2-\lambda^2 +i\epsilon]}u(P,S)\\ \nonumber
=\dfrac{1}{2p^+}\overline{u}(P',S')\Big[\dfrac{\Delta^{\perp}}{2M}G_1+\gamma^{\perp}(H+E+G_2)
+\dfrac{\Delta^{\perp}}{p^+}\gamma^+ G_3+\dfrac{i \epsilon_T^{\perp k}\Delta^{\perp}_k}{p^+}\gamma^+\gamma_5G_4\Big]{u}(P,S),
\end{eqnarray}

and the coefficient of the vector structure, $(H+E+G_2)$ is identified.  In Eq.(\ref {vectorQTM}) $g$ is the coupling strength, $m$ and  $\lambda$ are the quark and renormalization mass repectively.
To extract $G_2$ from the combination $(H+E+G_2)$,  $H$ is calculated  in QTM using the parametrization,

\begin{eqnarray}\label{H}
\dfrac{1}{2}\int{}\dfrac{dz^-}{2 \pi} e^{i x p^+z^-}\langle P',S'|\:\overline{q}(-\dfrac{z^-}{2}) \gamma^+ q (\dfrac{z^-}{2})\:|P,S\rangle
=\dfrac{1}{2p^+}\overline{u}(P',S')\Big[\gamma^+H+\dfrac{i \sigma^{+\rho}\Delta_{\rho}}{2M}E\Big]{u}(P,S).
\end{eqnarray}

For simplicity, only the divergent contributions are considered. $E$ is finite therefore does not contribute to the divergent parts of the matrix elements in Eqs. (\ref{vectorQTM}) and (\ref{H}). 

 To evaluate the $k^-$ integrals on the left-hand side of Eq.(\ref{vectorQTM}), the three regions regarding the interval of longitidunal momentum fraction ($x=k^+/ p^+$)  have to be distinguished. Identifying $\xi=\Delta^+/2p^+$ with the longitudinal momentum transfer fraction, 
 
\begin{itemize}
 \item  for  $\xi<x \leq 1,$  the incoming and outgoing longitudinal momentum fractions, $x-\xi$ and $x+\xi$ , are both positive and the correlator involves an incoming and outgoing quark.
 \item  For $-\xi \leq x\leq \xi,$ the incoming longitunal momentum fraction, $x-\xi$, is negative and outgoing, $x+\xi$, is positive. In this region the correlator involves an incoming anti-quark and outgoing quark.
\item For $-1\leq x< \xi$, both momentum fractions are negative, describing an incoming and outgoing anti-quark. 
  \end{itemize}  
The regions $\xi <x \leq 1$ and $-1\leq x< \xi$  are commonly referred to as DGLAP regions \cite{Gribov:1972ri,Lipatov:1974qm,Altarelli:1977zs,Dokshitzer:1977sg} and  $-1\leq x< \xi$ as ERBL region \cite{Efremov:1978rn,Lepage:1980fj}. 
 
Taking $m=\lambda=1$ for simplicity, the divergent part of $G_2$ in QTM is calculated as,

 \begin{equation}
    {G}_2=
    \begin{cases}
   \dfrac{g^2}{2\pi^2 }\dfrac{(1+x)}{(1-\xi^2)}\ln \Lambda_{\perp} & \text{for}\hspace{0.5 cm}\xi < x <1,\\
    -\dfrac{g^2}{4\pi^2}\dfrac{(1+x)}{\xi(1+\xi)}\ln \Lambda_{\perp} &  \text{for} \hspace{0.5 cm}-\xi \leq x\leq \xi, \\
      0 &  \text{for}\hspace{0.5 cm}-1< x<\xi,
    \end{cases}
 \end{equation}
where $\Lambda_{\perp}$ is the transverse momentum cut off. Since it violates the conservation of momentum in QTM, the distribution does not have a support in the region, $-1<  x< \xi$.

To calculate $\widetilde{G}_2$, the matrix element on the left-hand side of  Eq.(\ref{eq:axialvector}) is written using QTM with the vertex operator  $\Gamma=\gamma^{\perp}\gamma_5$ and the coefficient of the axial vector structure, $(\widetilde{H}+ \widetilde{G}_2)$ is identified,

\begin{eqnarray}\label{axialvectorQTM}
-\dfrac{ig^2}{2}\int{}\dfrac{d^4k}{(2 \pi)^4} \delta(k^+-xp^+)\overline{u}(P', S')\gamma^{\mu} \dfrac{(k\!\!\!/+\dfrac{\Delta\!\!\!/}{2}+m)}{[(k+\dfrac{\Delta}{2})^2-m^2+i \epsilon]} \gamma^{\perp} \gamma_5 \dfrac{(k\!\!\!/-\dfrac{\Delta\!\!\!/}{2}+m)}{[(k-\dfrac{\Delta}{2})^2-m^2+i\epsilon]}
  \gamma^{\nu}\\ \nonumber
\times \Big[g_{\mu\nu}  -\dfrac{n_{\nu}(p_{\mu}-k_{\mu})}{p^+-k^+}-\dfrac{n_{\mu}(p_{\nu}-k_{\nu})}{p^+-k^+}\Big] \dfrac{1 }{[(p-k)^2-\lambda^2 +i\epsilon]}u(P, S)\\  \nonumber
 =\dfrac{1}{2p^+}\overline{u}(P',S')\Big[\dfrac{\Delta^{\perp}}{2M}\gamma_5(\widetilde{E}+\widetilde{G}_1)+\gamma^{\perp}\gamma_5 (\widetilde{H}+\widetilde{G}_2)+\dfrac{\Delta^{\perp}}{p^+} \gamma^+\gamma_5\widetilde{G}_3+\dfrac{i \epsilon_T^{{\perp}k}\Delta^{\perp}_k}{p^+}\gamma^+\widetilde{G}_4\Big]{u}(P,S).
\end{eqnarray}

 $\widetilde{H}$ is calculated in the same model but with the vertex operator $\Gamma=\gamma^+\gamma_5$ using the parametrization,

\begin{eqnarray}\label{Htilde}
\dfrac{1}{2}\int{}\dfrac{dz^-}{2 \pi} e^{i x p^+z^-}\langle P',S'|\:\overline{q}(-\dfrac{z^-}{2}) \gamma^+\gamma_5  q (\dfrac{z^-}{2})\:|P,S\rangle
=\dfrac{1}{2p^+}\overline{u}(P',S')\Big[\gamma^+\gamma_5 \widetilde{H}+\dfrac{\gamma_5\Delta^+}{2M}\widetilde{E}\Big]\overline{u}(P,S).
\end{eqnarray}

Taking $m=\lambda=1$ for simplicity and considering only the divergent parts where $\widetilde{E}$  does not contribute to  the matrix elements in Eqs. (\ref{axialvectorQTM}) and (\ref{Htilde}), $\widetilde{G}_2$  is calculated as,

\begin{equation}
    \widetilde{G}_2=
    \begin{cases}
 \dfrac{g^2}{2\pi ^2}\dfrac{(x+\xi^2)}{(1-\xi^2)}\ln \Lambda_{\perp}& \text{for} \hspace{0.5 cm}\xi < x < 1,\\
    -\dfrac{g^2}{4 \pi^2}\dfrac{(x+\xi^2)}{\xi(1+\xi)} \ln \Lambda_{\perp} &  \text{for}\hspace{0.5 cm}-\xi \leq x\leq \xi, \\
      0 &  \text{for} \hspace{0.5 cm}-1< x< \xi,
    \end{cases}
 \end{equation}

\begin{figure}[ht]
\includegraphics[width=8.cm]{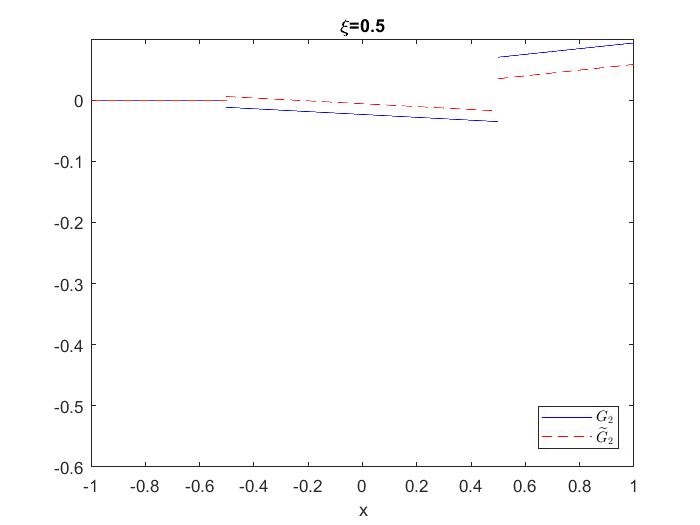}
\hspace{0.5cm}
\includegraphics[width=8.cm]{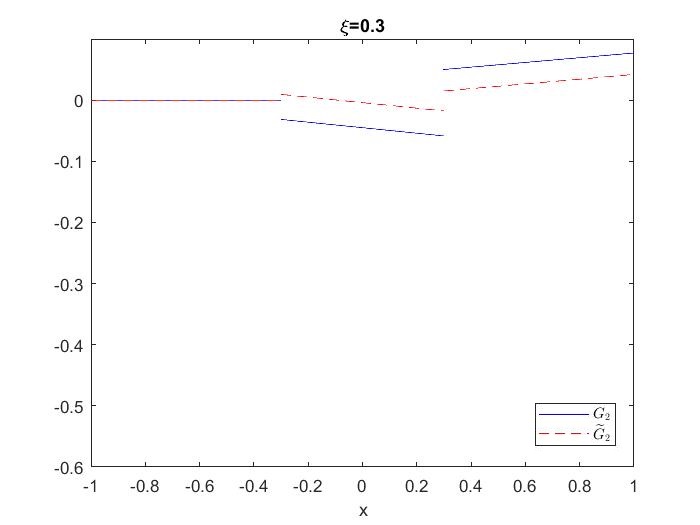}
\hspace{.5cm}
\includegraphics[width=8.cm]{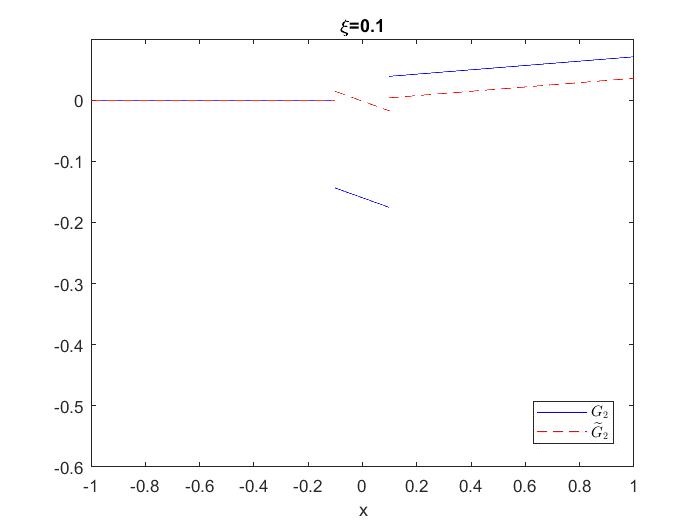}
\hspace{.5cm}
\includegraphics[width=8.cm]{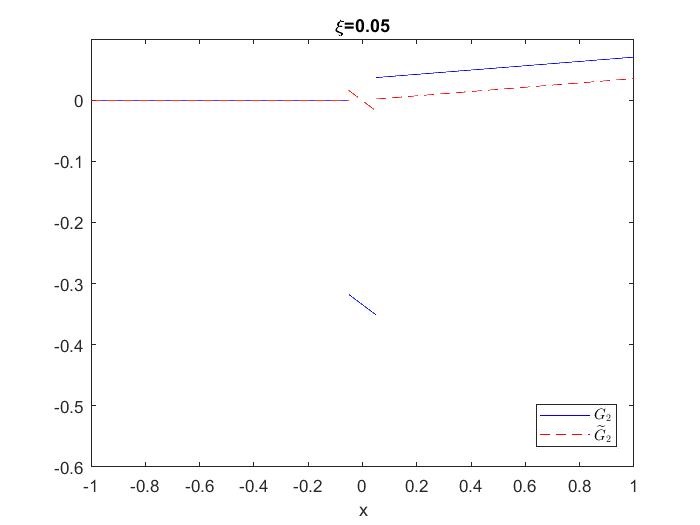}
\hspace{.5cm}
\includegraphics[width=8.cm]{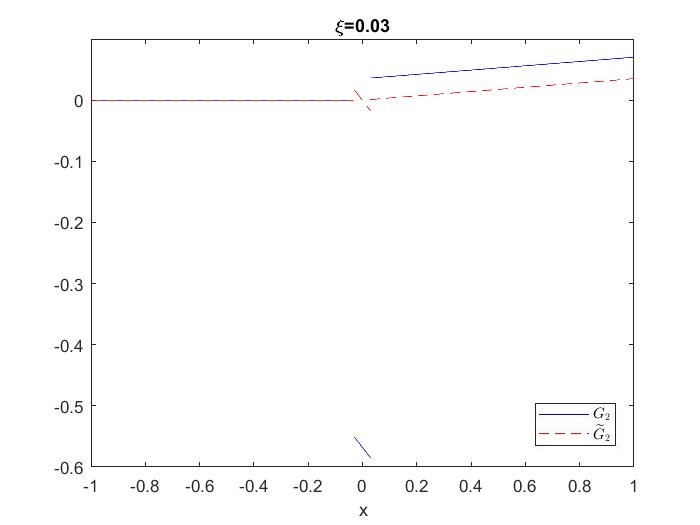}
\caption{Discontinuties of the twist-3 GPDs, $G_2$ and $\widetilde{G}_2$ in QTM for  $m=\lambda=1$, $\Lambda_{\perp}=2$,  and $g^2=1.$  }
\label{fig:GQTM}
\end{figure}

 As FIG. \ref{fig:GQTM} shows $G_2$ and $\widetilde{G}_2$  exhibit discontinuities at the points $x =\pm\xi$ which correspond to vanishing longitudinal momentum components in the initial or final state. In the limit of $\xi \rightarrow 0$, the discontinuities of  $\widetilde{G}_2$  stay finite and the contribution from the ERBL region partially cancels upon integrating over $x$. Whereas,  the discontinuities of $G_2$  diverge and its ERBL region resembles a representation of Dirac delta function as explained in the Appendix.
  In the following section the discontinuties of $G_2$ and $\widetilde{G}_2$ and their behaviors as $\xi\rightarrow0$ are investigated using SDM.

 \section{$G_2$ and $\widetilde{G}_2$ in Scalar Diquark Model}\label{section:SDM}
\begin{figure} [ht]\label{G_2(SDM)}
\centering
\includegraphics[width=8.cm]{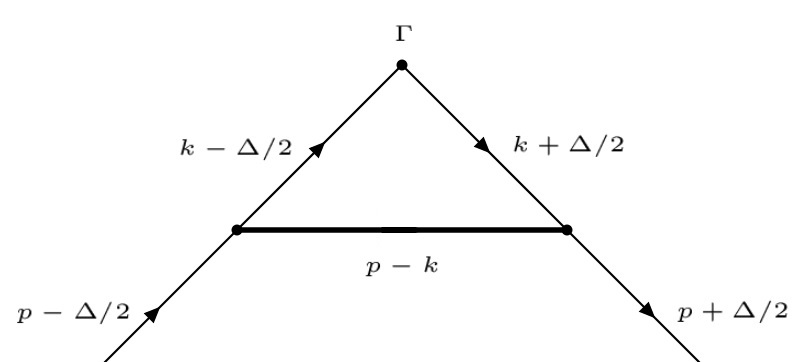}
\caption{Scalar diquark model  in a symmetric frame.}
\end{figure}

In scalar quark  diquark model (SDM) the three valence quarks of the nucleon are considered to be a bound state of a single quark and a scalar diquark.  We assume that the virtual  photon is interacting only with the single quark (active quark) and the  scalar diquark is only a spectator.

 To calculate $G_2$, the matrix element on the left-hand side of Eq.(\ref{eq:vector}) is written using SDM,

\begin{eqnarray}
\label{vectorSDM}
\dfrac{ig^2}{2}\int{}\dfrac{d^4k}{(2 \pi)^4} \delta(k^+-xp^+)\overline{u}(P',S')\dfrac{(k\!\!\!/+\dfrac{\Delta\!\!\!/}{2}+m)}{[(k+\dfrac{\Delta}{2})^2-m^2+i \epsilon]} \gamma^{\perp}  \dfrac{(k\!\!\!/-\dfrac{\Delta\!\!\!/}{2}+m)}{[(k-\dfrac{\Delta}{2})^2-m^2+i\epsilon]} 
\dfrac{1}{[(p-k)^2-\lambda^2 +i\epsilon]}u(P,S)\\ \nonumber
=\dfrac{1}{2p^+}\overline{u}(P',S')\Big[\dfrac{\Delta^{\perp}}{2M}G_1+\gamma^{\perp} (H+E+G_2)+\dfrac{\Delta^{\perp}}{p^+}\gamma^+ G_3+\dfrac{i \epsilon_T^{{\perp}k}\Delta^{\perp}_k}{p^+}\gamma^+\gamma_5G_4\Big]{u}(P,S),
\end{eqnarray}
and the coefficient of the vector structure, $(H+E+G_2)$, is identified. In Eq.(\ref{vectorSDM}) $m$ and $\lambda$ denotes the quark and diquark mass respectively. $G_2$ is extracted considering only the divergent contributions as in the QTM case,

\begin{equation}
    {G}_2=
    \begin{cases}
  -\dfrac{g^2}{4\pi^2}\dfrac{(1-x)}{(1-\xi^2)} \ln \Lambda_{\perp} & \text{for} \hspace{0.5 cm}\xi < x<1,\\
  -  \dfrac{g^2}{16 \pi^2}  \dfrac{(2x+\xi-1)}{ \xi (1+\xi)}  \ln \Lambda_{\perp}&  \text{for} \hspace{0.5 cm}-\xi \leq x\leq \xi, \\
      0 &  \text{for} \hspace{0.5 cm}-1< x< \xi.
    \end{cases}
 \end{equation}
 
\begin{figure}[ht]
\includegraphics[width=8.cm]{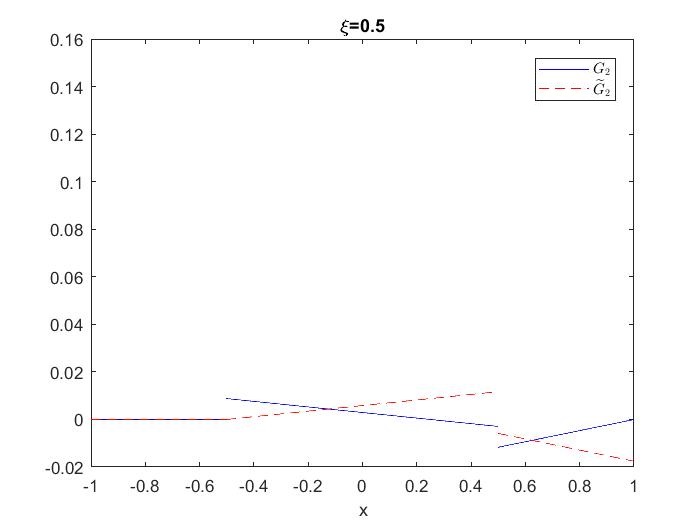}
\hspace{.5cm}
\includegraphics[width=8.cm]{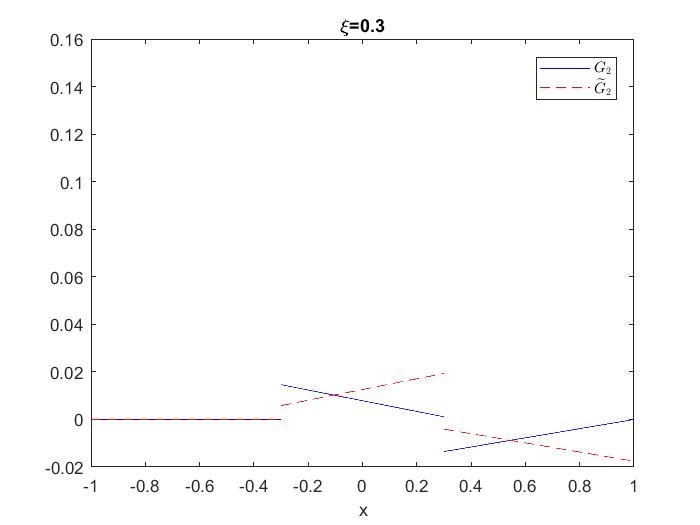}
\hspace{.5cm}
\includegraphics[width=8.cm]{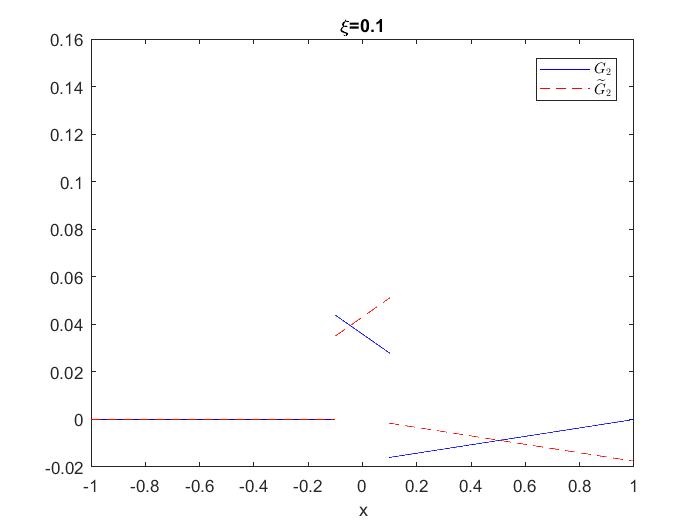}
\hspace{.5cm}
\includegraphics[width=8.cm]{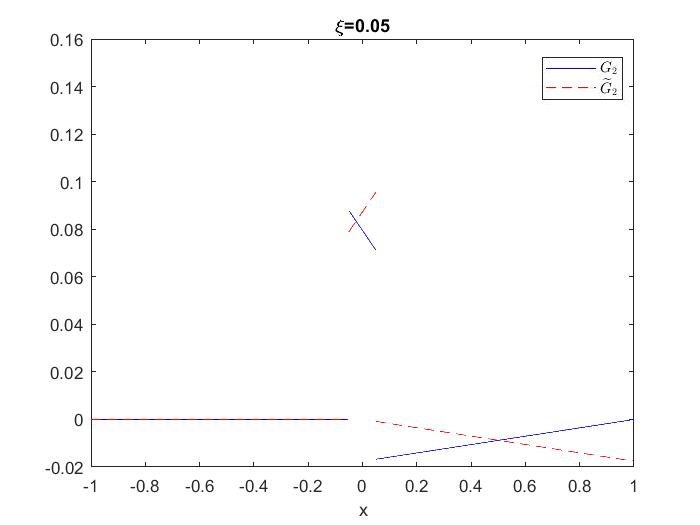}
\hspace{.5cm}
\includegraphics[width=8.cm]{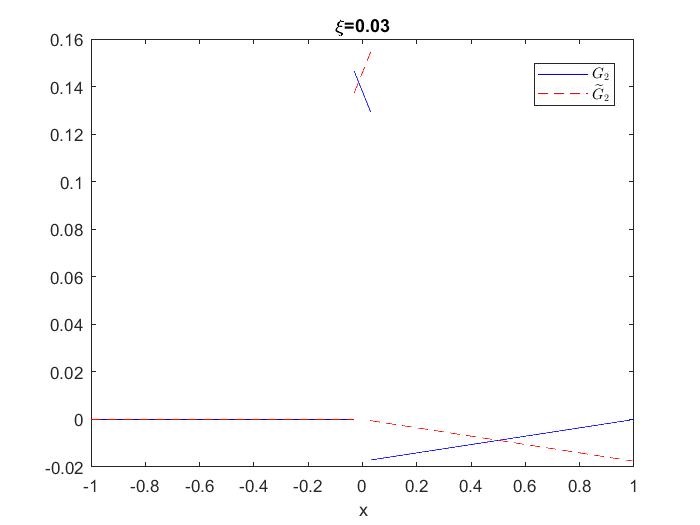}
\caption{Discontinuties of the twist-3 GPDs, $G_2$ and $\widetilde{G}_2$ in SDM for $M=m=\lambda=1$, $\Lambda_{\perp}=2$ and $g=1$.}
\label{fig:SDM}
\end{figure}

To calculate $\widetilde{G}_2$  the matrix element in Eq.(\ref{eq:axialvector}) is written using SDM,
\begin{eqnarray}\label{axialvectorSDM}
\dfrac{ig^2}{2}\int{}\dfrac{d^4k}{(2 \pi)^4} \delta(k^+-xp^+)\overline{u}(P',S')\dfrac{(k\!\!\!/+\dfrac{\Delta\!\!\!/}{2}+m)}{[(k+\dfrac{\Delta}{2})^2-m^2+i \epsilon]} \gamma^{\perp} \gamma_5 \dfrac{(k\!\!\!/-\dfrac{\Delta\!\!\!/}{2}+m)}{[(k-\dfrac{\Delta}{2})^2-m^2+i\epsilon]} 
\dfrac{1}{[(p-k)^2-\lambda^2 +i\epsilon]}u(P,S)\\  \nonumber
=\dfrac{1}{2p^+}\overline{u}(P',S')\Big[\dfrac{\Delta^{\perp}}{2M}\gamma_5(\widetilde{E}+\widetilde{G}_1)+\gamma^{\perp}\gamma_5 (\widetilde{H}+\widetilde{G}_2)+\dfrac{\Delta^{\perp}}{p^+} \gamma^+\gamma_5\widetilde{G}_3+\dfrac{i \epsilon_T^{{\perp}k}\Delta^{\perp}_k}{p^+}\gamma^+\widetilde{G}_4\Big]{u}(P,S),
\end{eqnarray}
the axial vector structure coefficient, $(\widetilde{H}+\widetilde{G}_2)$ is identified
and the divergent part of $\widetilde{G}_2$ is extracted, 
 \begin{equation}
    \widetilde{G}_2=
    \begin{cases}
    \dfrac{g^2}{4\pi^2} \dfrac{(\xi^2-x)}{(1-\xi^2)}\ln \Lambda_{\bot} & \text{for}\hspace{0.5 cm}\xi < x <1,\\
     \dfrac{g^2}{16 \pi^2 } \dfrac{(2x-2\xi^2+\xi+1)}{\xi(1+\xi)} \ln \Lambda_{\bot}&  \text{for} \hspace{0.5 cm}-\xi \leq x\leq \xi, \\
      0 &  \text{for} \hspace{0.5 cm}-1< x< \xi.
    \end{cases}
 \end{equation}
As shown in FIG. \ref{fig:SDM},  $G_2$ and $\widetilde{G}_2$  exhibit discontinuities at the points $x =\pm\xi$ also in SDM. These discontinuities  diverge as $\xi \rightarrow 0$ and the ERBL regions  of both GPDs resemble a representation of a Dirac delta function. 
 
The behavior of the discontinuties of   $\widetilde{G}_2$  and $G_2$  as $\xi\rightarrow 0$ in QTM and SDM are summarized in TABLE \ref{tab:discontinuties}.   In the following section twist-3 PDFs and quasi-PDFs are investigated using SDM.

 \begin{table}[ht]
 \centering
  \begin{tabular}{c || c | c}
  \hline \\
 Twist-3 GPD &  QTM  & SDM\\ \hline      \hline\\ 
 $G_2$ & Divergent &  Divergent \\ \hline \\ 
$\widetilde{G}_2$ & Finite  & Divergent   \\ \hline 
    \end{tabular}
    \caption{The behavior of the discontinuties of the twist-3 GPDs, $\widetilde{G}_2$  and $G_2$  as $\xi\rightarrow 0$ in QTM and SDM.}
    \label{tab:discontinuties}
\end{table}


\section{ Twist-3 PDFs and Twist-3 quasi-PDFs in Scalar Diquark Model}\label{section:pdf}

For equal momenta and spins of the initial and final hadron states, the matrix elements defining GPDs reduce to the matrix elements defining PDFs \cite{Collins:1981uw}. There are three quark PDFs  $(f_1, g_1, h_1)$  at twist-2 level and three quark PDFs $(e, h_L, g_2)$  at twist-3 level. The complete set of PDFs is defined by the matrix elements of quark bilocal operators,

\begin{eqnarray}\label{f1}
\int{}\dfrac{d\lambda}{2\pi}e^{i \lambda x}\langle P,S|\overline{q}(0)\gamma^{\mu}q(\lambda n)|P,S\rangle=2\big[f_1(x) \hat{p}^{\mu}+ M^2 f_4(x) \hat{n}^{\mu} \big],
\end{eqnarray}
\begin{eqnarray}\label{gT}
\int{}\dfrac{d\lambda}{2\pi}e^{i \lambda x}\langle P,S|\overline{q}(0)\gamma^{\mu}\gamma_5q(\lambda n)|P,S\rangle=2\big\{g_1(x) \hat{p}^{\mu}(S \cdot \hat{n})+g_T(x)S_{\bot}^{\mu}+ M^2 g_3(x) \hat{n}^{\mu}(S \cdot \hat{n}) \big\},
\end{eqnarray}
\begin{eqnarray}\label{e}
\int{}\dfrac{d\lambda}{2\pi}e^{i \lambda x}\langle P,S|\overline{q}(0)q(\lambda n)|P,S\rangle=2Me(x),
\end{eqnarray}
\begin{eqnarray}\label{hL}
\int{}\dfrac{d\lambda}{2\pi}e^{i \lambda x}\langle P,S|\overline{q}(0)i\sigma^{\mu\nu}\gamma_5q(\lambda n)|P,S\rangle=2\big[h_1(x)\dfrac{(S^{\mu}_{\perp}\hat{p}^{\nu}-S^{\nu}_{\perp}\hat{p}^{\mu})}{M}+h_L(x) M(\hat{p}^{\mu}\hat{n}^{\nu}-\hat{p}^{\nu}\hat{n}^{\mu}) S\cdot\hat{ n} \\ \nonumber
+h_3(x)M(S_{\perp}^{\mu}\hat{n}^{\nu}-S_{\perp}^{\nu}\hat{n}^{\mu})\big],
\end{eqnarray}
where P, S and M are the momentum, spin and mass of the parent hadron respectively. $\hat{p}$ and $\hat{n}$ are  light-like vectors, i.e. $\hat{p}^2=\hat{n}^2=0$, with the components, $\hat{p}^-=\hat{p}_{\perp}=0$, $\hat{p}^+=P^+$ and $\hat{n}^+=\hat{n}_{\perp}=0 , \hat{n}^-=1/P^+$. The spin vector, $S^{\mu}$ is decomposed as $S^{\mu}=(S\cdot \hat{n}) \hat{p}^{\mu}+(S\cdot \hat{p}) \hat{n}^{\mu}+S_{\perp}^{\mu}.$ $x$ represents the parton's light-cone momentum fraction and each PDF has a support in the interval, $-1\leq x \leq 1$. The PDFs $f_4, g_3$ and $h_3$ appear at twist-4 level. In the parametrization given by Eq.(\ref{gT}) $g_T(x)=g_1(x)+g_2(x)$. 

Even though the matrix elements entering the cross section are usually dominated by twist-2 operators in the Bjorken limit,  and twist-3 operators are mostly relevant for subleading corrections, the twist-3 PDFs, $g_2(x)$ and $h_L(x)$ are unique in the sense that they appear as leading contributions in some spin asymmetries; for example, $g_2(x)$ can be measured in the transversely polarized DIS and $h_L(x)$ can be measured in the longitudinal-transverse double spin asymmetry in the polarized Drell-Yan process \cite{Jaffe:1991ra}. 

The twist-3 GPD, $\widetilde{G}_2$ reduces to the twist-3 PDF, $g_2(x)$ in the forward limit. In order to investigate the Dirac delta function behavior of $\widetilde{G}_2$ in the ERBL region in the forward limit, $g_2(x)=g_T(x)-g_1(x)$ is calculated as explained in Appendix.

\begin{figure}[ht ]\label{g2(SDM)}
\centering
\includegraphics[width=8.cm] {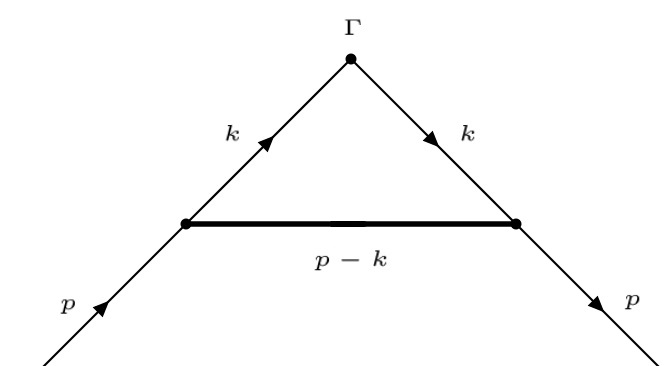}
\caption{Scalar diquark model in the forward limit.}
\end{figure}

$g_2(x)$ contains the following term which involves  a Dirac delta function, $\delta(x)$,
\begin{eqnarray}
g_{2,\delta}(x)=\dfrac{g^2}{16 \pi^2}\dfrac{(x+\dfrac{m}{M})}{(1-x)} \ln(\dfrac{\Lambda^2_{\perp}+m^2}{m^2})\delta(x),
\end{eqnarray}
where $m$ is the quark mass, $M$ is the nucleon mass and $\Lambda_{\perp}$ is the transverse momentum cut off.

$g_{2,\delta}(x)$  originates from the light-cone energy integral,
\begin{eqnarray}\label{originofthedelta}
\int{}\dfrac{dk^-}{(k^2-m^2+i\epsilon)^2}
\end{eqnarray}
which is performed using Cauchy's theorem.

Obviously for $k^+\neq0$, 
\begin{eqnarray}\label{kminusint1}
\int{}\dfrac{dk^-}{(k^2-m^2+i\epsilon)^2}=\int{}\dfrac{dk^-}{\Big[2k^+\Big(k^--\dfrac{(k_{\perp}^2+m^2)}{2k^+}+\dfrac{i\epsilon}{2k^+}\Big)\Big]^2}=0,
\end{eqnarray}
since enclosing the double pole, $k^-=\dfrac{(k_{\perp}^2+m^2)}{2k^+}-\dfrac{i\epsilon}{2k^+}$ can be avoided by closing the contour in approriate half-plane of the complex $k^-$-plane. However,

\begin{eqnarray}\label{kminusint2}
\int{}dk^+dk^-\dfrac{1}{(k^2-m^2+i\epsilon)^2}&=\int{}dk^+dk^-\dfrac{1}{(2k^+k^--k_{\perp}^2-m^2+i\epsilon)^2}\\ \nonumber
&=\int{}d^2k_L\dfrac{1}{(k_L^2-k_{\perp}^2-m^2+i\epsilon)^2}=\dfrac{i\pi}{k_{\perp}^2+m^2}.
\end{eqnarray}
Combining Eq.(\ref{kminusint1}) with Eq.(\ref{kminusint2}) thus implies,
\begin{eqnarray}
\int{}\dfrac{dk^-}{(k^2-m^2+i\epsilon)^2}=\dfrac{i\pi}{k_{\perp}^2+m^2}\delta(k^+).
\end{eqnarray}

Therefore the singular term is,
\begin{eqnarray}\label{g2deltak}
g_{2,\delta}(k^+)=-\dfrac{ig^2}{(2\pi)^2}\dfrac{(k^++\dfrac{m}{M}P^+)}{(P^+-k^+)}\int{}\dfrac{d^2k_{\perp}}{(2\pi)^2} \dfrac{dk^-}{(k^2-m^2+i\epsilon)^2}=\dfrac{g^2}{16 \pi^2}\dfrac{(k^++\dfrac{m}{M}P^+)}{(P^+-k^+)} \ln(\dfrac{\Lambda^2_{\perp}+m^2}{m^2})\delta(k^+)
\end{eqnarray}

The term in Eq.(\ref{g2deltak}) is given in terms of longitudinal momentum, $k^+$. In order to express it as a function of longitudinal momentum fraction, it is integrated over $k^+$ and multiplied by $P^+$,

\begin{eqnarray}\label{g2delta}
g_{2,\delta}(x)=\dfrac{g^2}{16 \pi^2}\dfrac{(x+\dfrac{m}{M})}{(1-x)} \ln(\dfrac{\Lambda^2_{\perp}+m^2}{m^2})\delta(x).
\end{eqnarray}

 We now consider the twist-3 quasi-PDF, $g_2^{quasi}(x)$  to illustrate the origin of the Dirac delta function contribution. While PDFs are calculated using light-cone coordinates, when calculating quasi-PDFs one treats the operators in normal coordinates where the nucleon moves purely in spatial direction with a momentum $P^z$ \cite{ Ji:2013dva,Ji:2014gla}. PDFs are recovered from quasi-PDFs by taking the limit $P^z\rightarrow \infty$ \cite{Xiong:2013bka,Stewart:2017tvs,Izubuchi:2018srq}.

\begin{figure}[ht]
\includegraphics[width=8.cm]{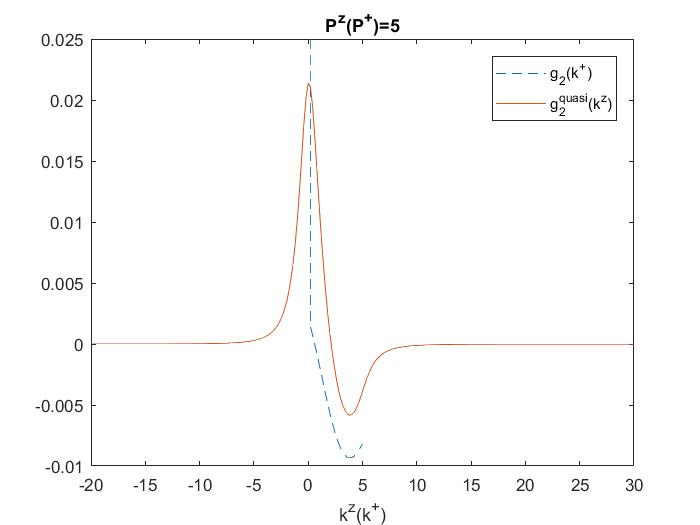}
\hspace{0.5cm}
\includegraphics[width=8.cm]{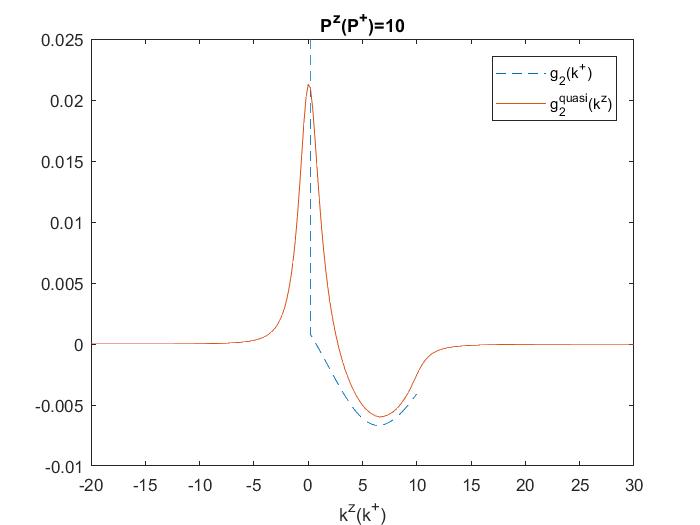}
\hspace{.5cm}
\includegraphics[width=8.cm]{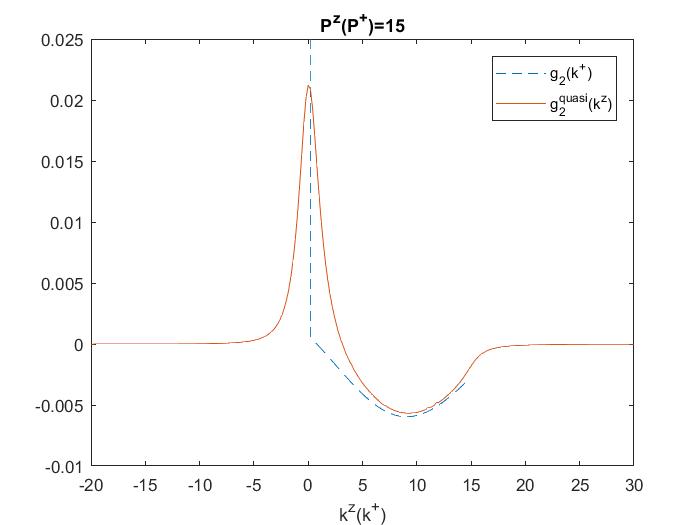}
\hspace{.5cm}
\includegraphics[width=8.cm]{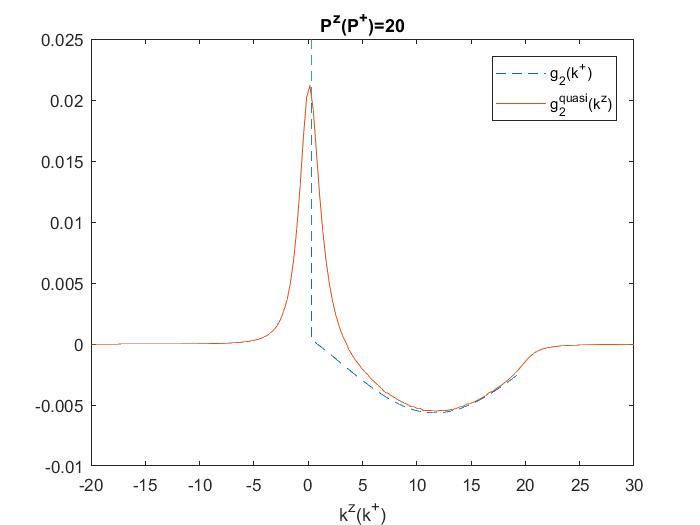}
\hspace{.5cm}
\includegraphics[width=8.cm]{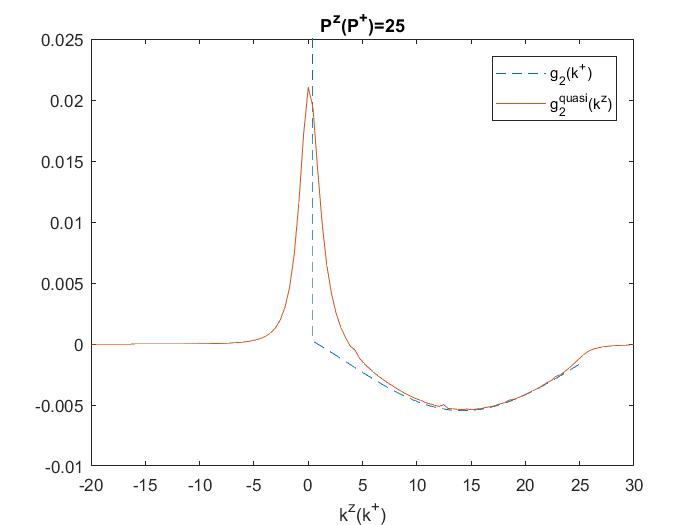}
\caption{$g_2(k^+)$ and $g_2^{quasi}(k^z)$ in SDM for $m=M=\lambda=1$ and   $\Lambda_{\perp}=2$.  }
\label{fig:g2(k)}
\end{figure}

In addition, we also consider the distributions as functions of  longitudinal momenta, i.e. $g_2(k^+)$ and $g_2^{quasi}(k^z)$ as shown in FIG.\ref{fig:g2(k)}.  The twist-3 distributions are identified by their scaling property under a  longitudinal nucleon momentum boost, i.e twist-3 PDFs scale with $1/{P^+}$ and twist-3 quasi-PDFs with  $1/{P^z}$.
Whereas the distributions in FIG.\ref{fig:g2(k)} have two components one of which obey the twist-3 scaling properties while the other component at $k^z(k^+)=0$ does not scale as the nucleon is boosted to higher longitudinal momenta, i.e. does not change as $1/P^z (1/P^+)$.

\begin{figure}[ht]
\includegraphics[width=8.cm]{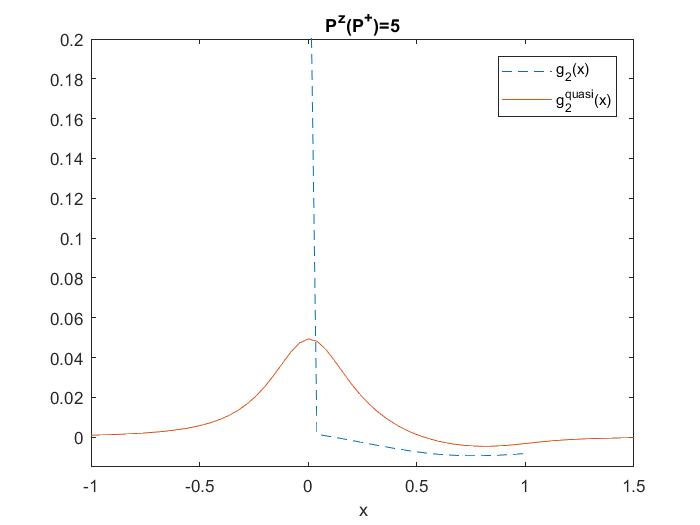}
\hspace{0.5cm}
\includegraphics[width=8.cm]{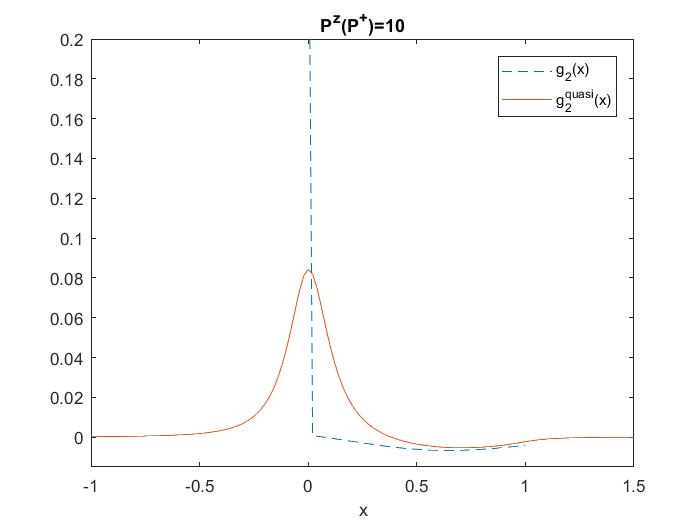}
\hspace{.5cm}
\includegraphics[width=8.cm]{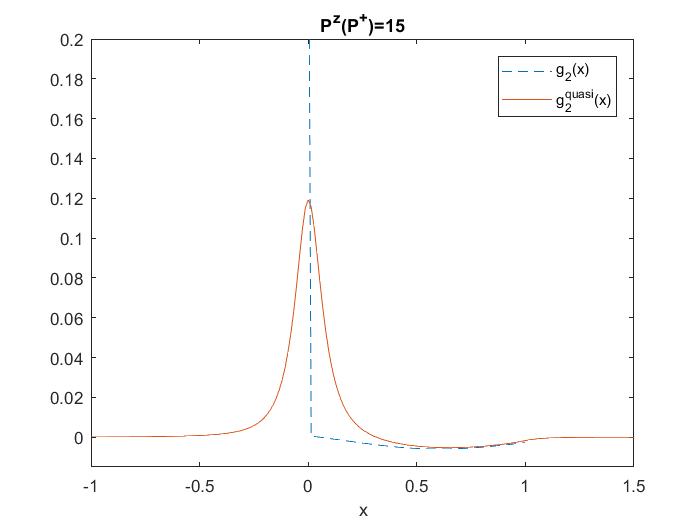}
\hspace{.5cm}
\includegraphics[width=8.cm]{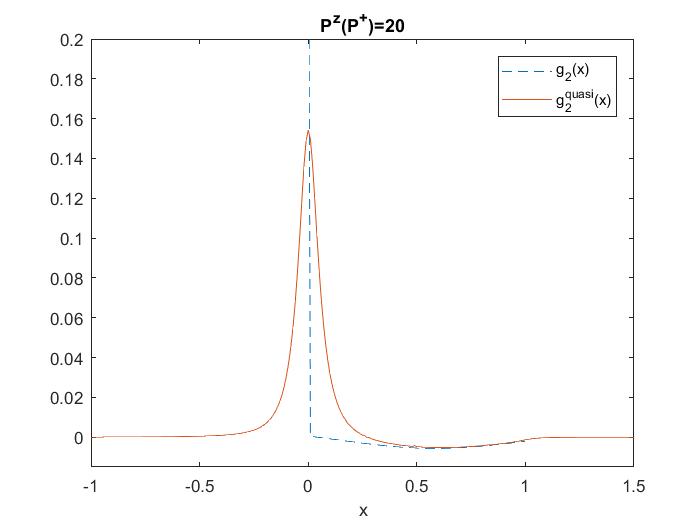}
\hspace{.5cm}
\includegraphics[width=8.cm]{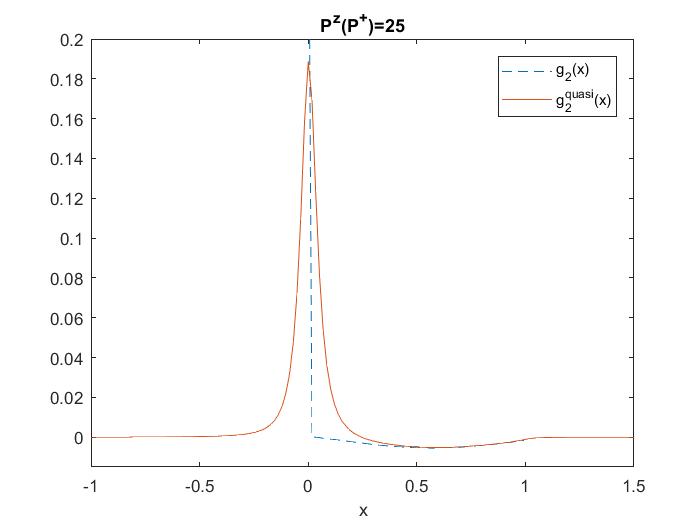}
\caption{$g_2(x)$ and $g_2^{quasi}(x)$ in SDM for $m=M=\lambda=1$ and  $\Lambda_{\perp}=2$. $x=\dfrac{k^+}{P^+}$ for $g_2(x)$ and $x=\dfrac{k^z}{P^z}$ for $g_2^{quasi}(x)$.}
\label{fig:px}
\end{figure}

The non-scaling component of  PDF, $g_2(k^+)$ corresponds to the term  $g_{2,\delta}(k^+)$ given by Eq.(\ref{g2deltak}) while  the non-scaling component of  the quasi-PDF $g_2^{quasi}(k^z)$ originates from the term,
\begin{eqnarray}\label{g2nsquasik}
g_{2,ns}^{quasi}(k^z)=-\dfrac{g^2}{16 \pi^2}\dfrac{(k^z+\dfrac{m}{M}P^z)}{(P^z-k^z)}\dfrac{1}{(k_{\perp}^2+k_z^2+m^2)^{1/2}}\Big|_{k_{\perp}=0}^{\Lambda_{\perp}}.
\end{eqnarray}

 FIG.\ref{fig:px} shows the distributions  as  functions of $x$ where the expression in Eq.(\ref{g2nsquasik}) is multiplied by $P^z$. It can be seen that the non-scaling component of the quasi-PDF  $g_2^{quasi} (x)$ grows proportionally to the longitudinal momentum boost  while its width decreases and has support only for $x=0$. Therefore as the nucleon is boosted to the IMF, it can be identified with a representation of a Dirac delta function at $x=0$ .
\begin{eqnarray}
g_{2,ns}^{quasi}(x)=lim_{P^z\rightarrow\infty}-\dfrac{g^2}{16 \pi^2}\dfrac{(x+\dfrac{m}{M})}{(1-x)}\dfrac{P^z}{\big[k_{\perp}^2+(xP^z)^2+m^2\big]^{1/2}}\Big|_{k_{\perp}=0}^{\Lambda_{\perp}}=\delta(x)
\end{eqnarray}

Operator product expansion (OPE) analysis of the matrix elements allows the  twist-3 distributions to be decomposed into the contributions expressed in terms of twist-2 distributions (WW-part) \cite {Wandzura:1977qf}, a quark mass term and the rest which involves interactions. For example for $g_2(x)$ the decomposition reads,
  \begin{eqnarray}\label{decomposition}
  g_2(x)=g_2^{WW}(x)+g_2^m(x)+g_2^3(x)
  \end{eqnarray}
  where the WW-part is given by  \cite{Harindranath:1997qn},
   \begin{eqnarray}\label{g2WW}
  g_2^{WW}(x)=-g_1(x)+\int_x^1{\dfrac{dy}{y} g_1(y)}.
  \end{eqnarray}
In QCD  $g_2^3(x)$ corresponds to the pure quark-gluon correlation part of the twist-3 distribution. These pure quark-gluon correlations which are also called \textit{genuine twist-3} terms, involve a novel type of information  that  is not contained in  twist-2 distributions. For example, the $x^2$ moment of the genuine twist-3 part of  polarized PDF $g_2(x)$ can be identified with the transverse component of the color-Lorentz force acting on the struck quark at the instant after absorbing the virtual photon \cite{Burkardt:2008ps}.
 
  An important question is whether the components which do not scale under a longitudinal momentum boost come from the WW parts of  twist-3 distributions. In order to address this question, in addition to $g_1(x)$ and $g_2(x)$, also the twist-2 PDFs $f_1(x), h_1(x)$ and twist-3 pdfs $e(x), h_L(x)$  are calculated using SDM. The PDFs calculated with  QTM are taken from Ref.\cite{Burkardt:2001iy}.

\begin{table}[ht]
 \centering
  \begin{tabular}{ c | c | c}
    \hline\\
    Twist-2 PDF& SDM& QTM \\ \hline
        \hline
    $f_1(x)$ & $\times$  &  $\times$  \\ \hline

   $ g_1(x)$ &$\times$   & $\times$   \\ \hline
   $h_1(x)$  & $\times$  & $\times$   \\ 
    \hline
  \end{tabular}
  
  \vspace{0.5cm}
  \begin{tabular}{ c | c | c }
    \hline\\
    Twist-3 PDF & SDM& QTM\\  \hline
        \hline
    $e(x)$ & $\surd$  &  $\surd$  \\ \hline
 $ h_L(x)$ &$\surd$    & $\surd$   \\ \hline
   $g_2(x)$  & $\surd$   & $\times$   \\ 
    \hline
    \end{tabular}
      \caption{Dirac delta functions in PDFs calculated using SDM and QTM.$''\times''$ denotes there is no $\delta(x)$ and $''\surd''$ denotes there is a $\delta(x)$ in a PDF.}
        \label{table:deltafunctions}
  \end{table}

\begin{figure}[ht]
\includegraphics[width=8.cm]{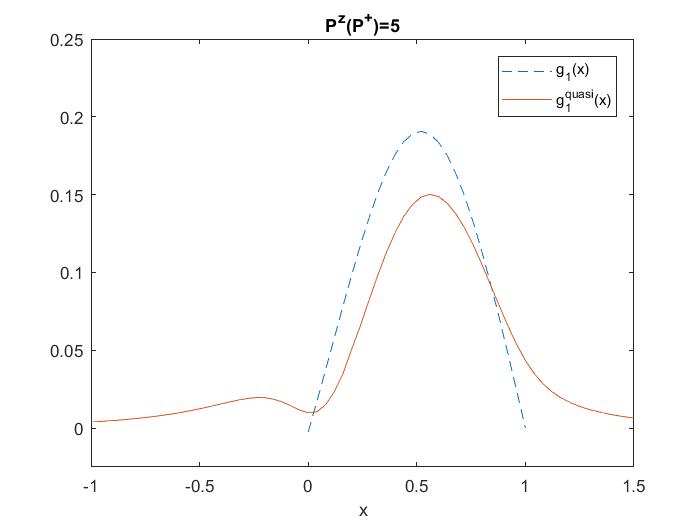}
\hspace{0.5cm}
\includegraphics[width=8.cm]{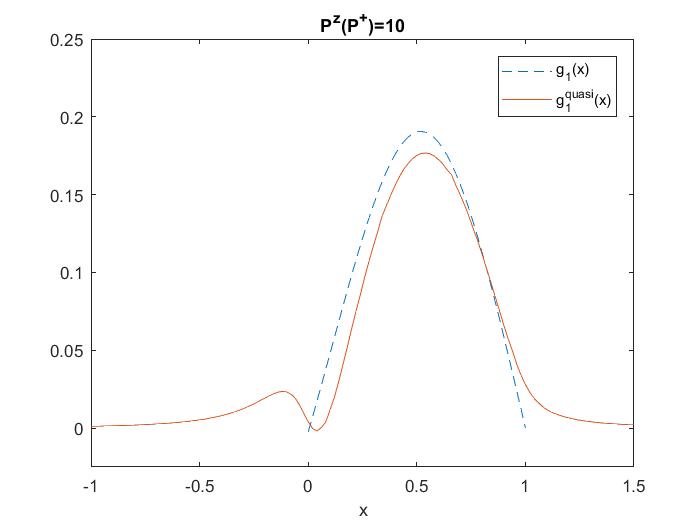}
\hspace{.5cm}
\includegraphics[width=8.cm]{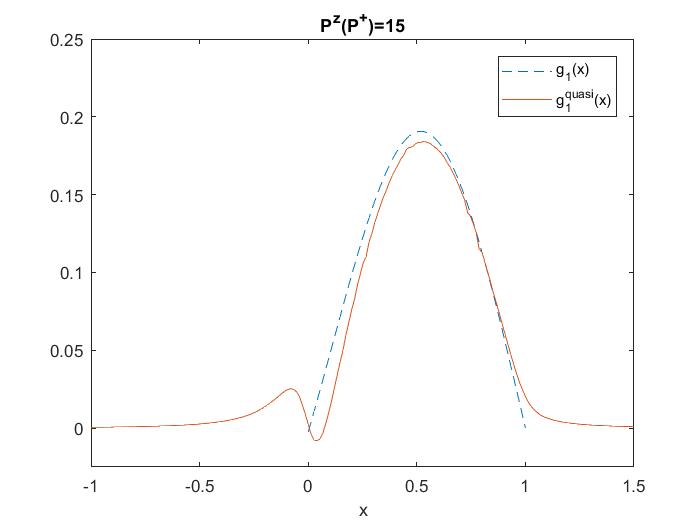}
\hspace{.5cm}
\includegraphics[width=8.cm]{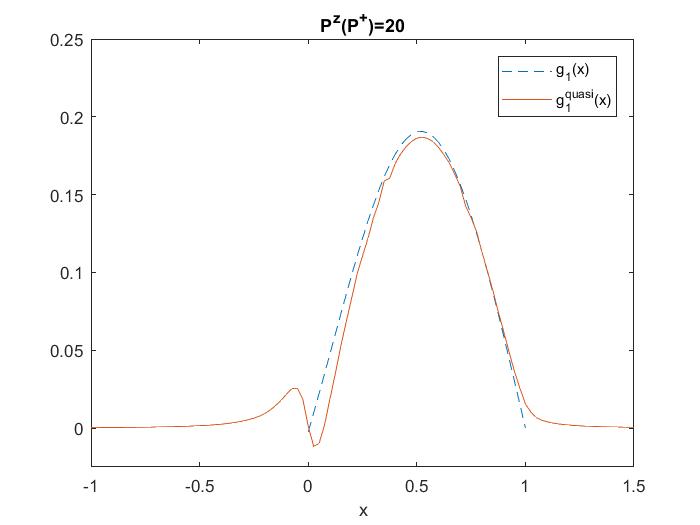}
\hspace{.5cm}
\includegraphics[width=8.cm]{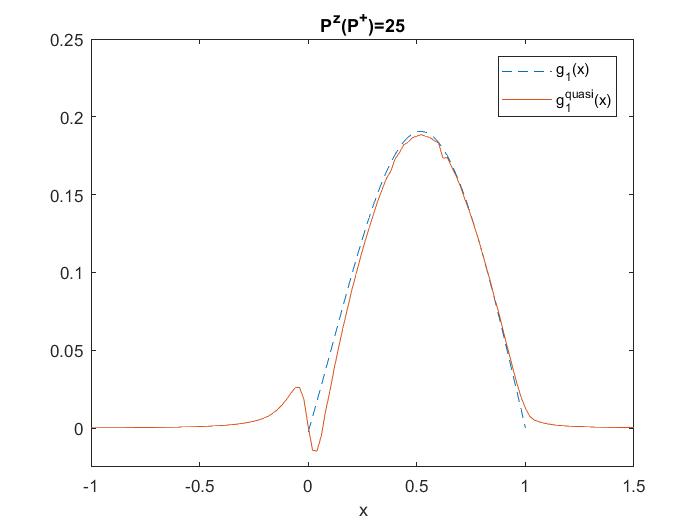}
\caption{$g_1(x)$ and $g_1^{quasi}(x)$ in  in SDM for $m=M=\lambda=1$ and $\Lambda_{\perp}=2$. $x=\dfrac{k^+}{P^+}$ for $g_1(x)$ and $x=\dfrac{k^z}{P^z}$ for $g_1^{quasi}(x)$.} 
\label{g1(x)}
\end{figure}

\begin{figure}[ht]
\includegraphics[width=8.cm]{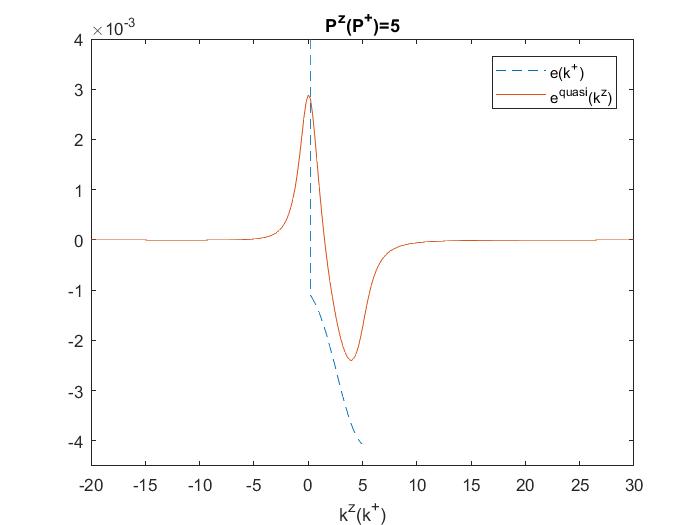}
\hspace{0.5cm}
\includegraphics[width=8.cm]{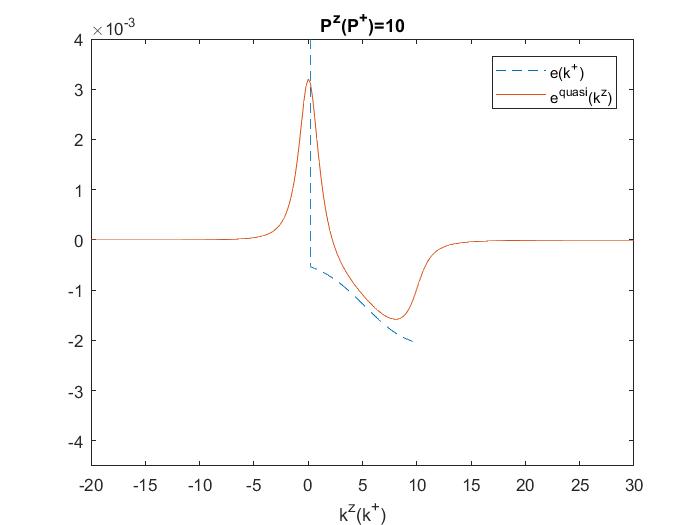}
\hspace{.5cm}
\includegraphics[width=8.cm]{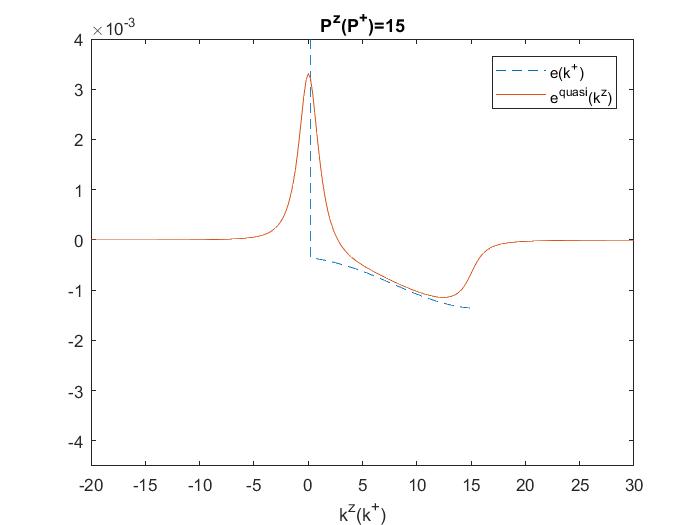}
\hspace{.5cm}
\includegraphics[width=8.cm]{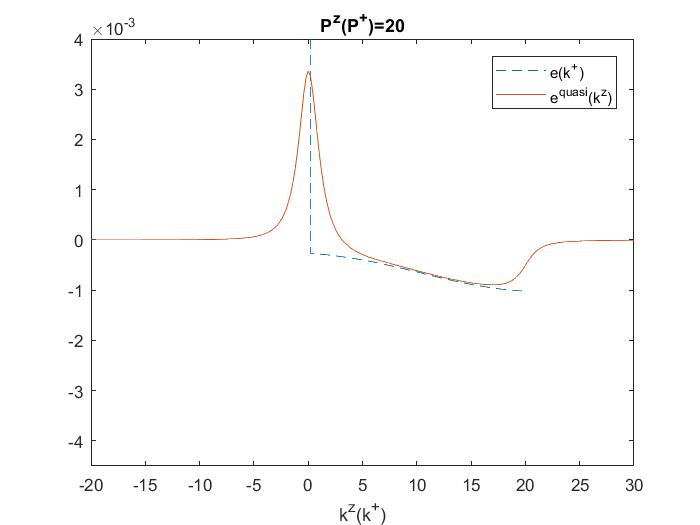}
\hspace{.5cm}
\includegraphics[width=8.cm]{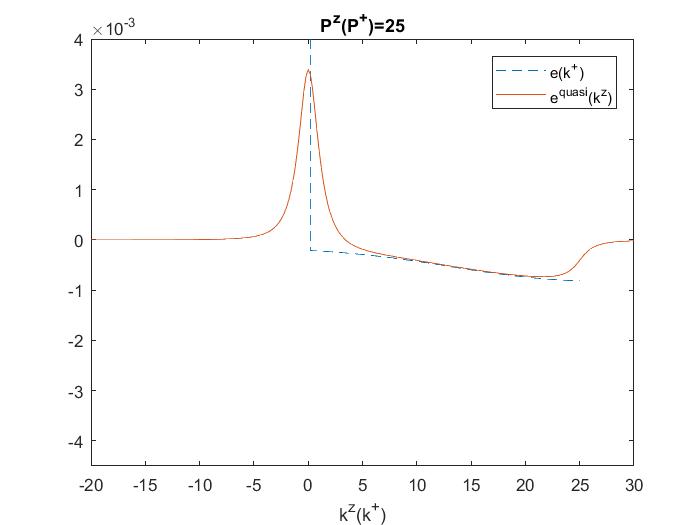}
\caption{$e(k^+)$ and $e^{quasi}(k^z)$ in in SDM for $m=M=\lambda=1$ and  $\Lambda_{\perp}=2$. }
\label{fig:ek}
\end{figure}

\begin{figure}[ht]
\includegraphics[width=8.cm]{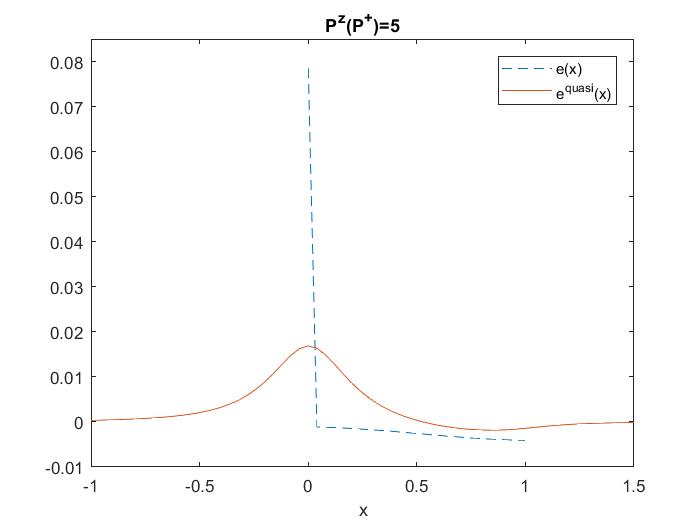}
\hspace{0.5cm}
\includegraphics[width=8.cm]{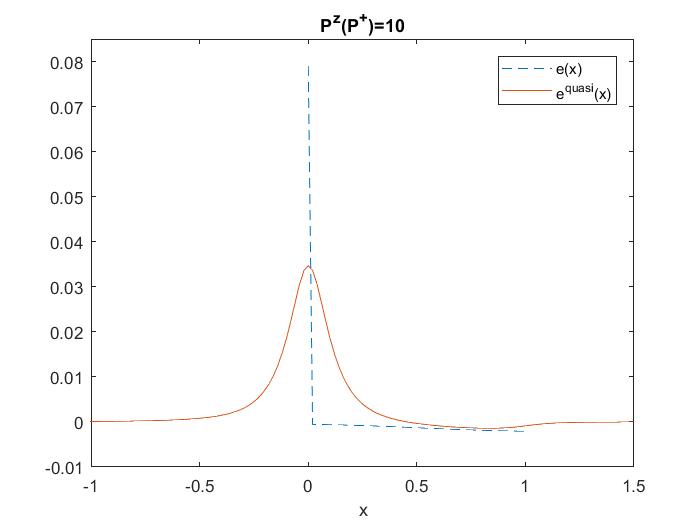}
\hspace{.5cm}
\includegraphics[width=8.cm]{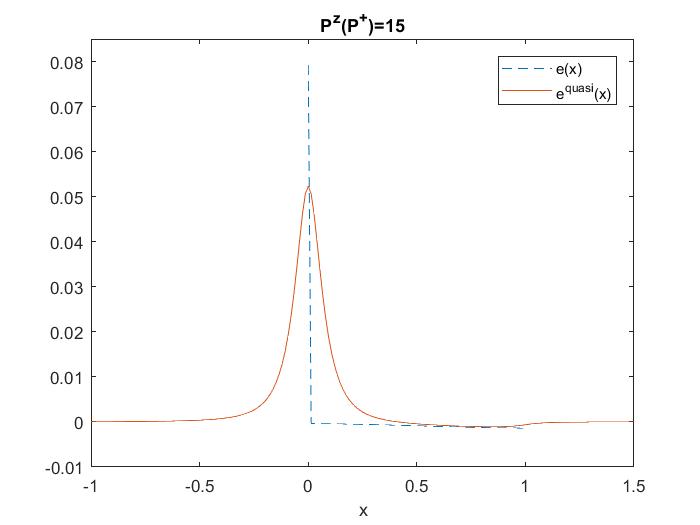}
\hspace{.5cm}
\includegraphics[width=8.cm]{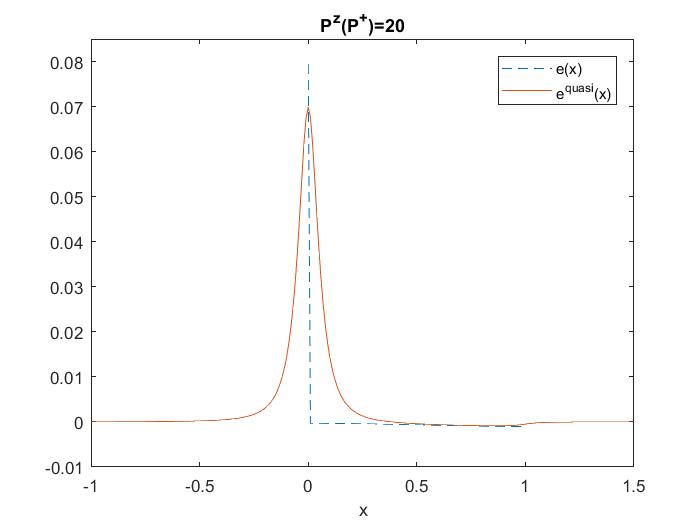}
\hspace{.5cm}
\includegraphics[width=8.cm]{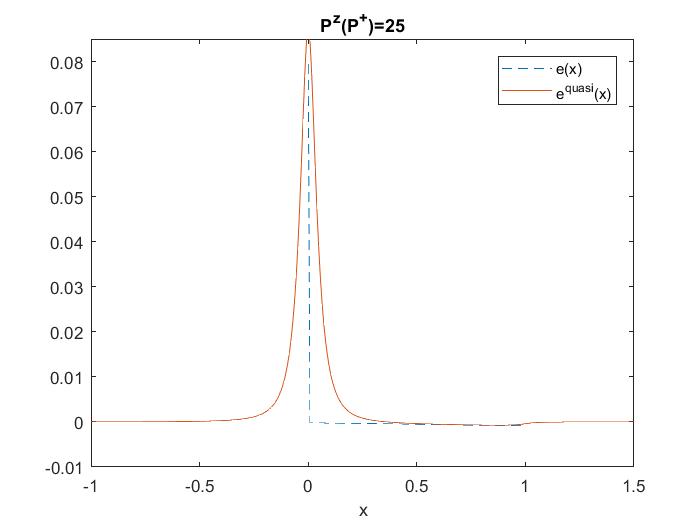}
\caption{$e(x)$ and $e^{quasi}(x)$ in  in SDM for $m=M=\lambda=1$ and  $\Lambda_{\perp}=2$. $x=\dfrac{k^+}{P^+}$ for $e(x)$ and $x=\dfrac{k^z}{P^z}$ for $e^{quasi}(x)$.} 
\label{fig:ex}
\end{figure}

As shown in TABLE \ref{table:deltafunctions}, all the  twist-3 PDFs calculated in SDM and QTM contain a $\delta(x)$ term only with the exception of $g_2(x)$ in QTM. Whereas, such a term does not appear in any of the twist-2 PDFs. As an example, the twist-2 PDF  $g_1(x)$ and the  twist-2 quasi-PDF $g_1^{quasi}(x)$ are calculated in SDM. FIG. \ref{g1(x)} shows that  $g_1^{quasi}(x)$ converges to $g_1(x)$ as $ P^z\rightarrow\infty$    without generating a $\delta(x)$. Therefore any potential singularity which can result from the integral in Eq.(\ref{g2WW}) does not originate from Eq.(\ref{originofthedelta}). For this reason the WW part is not the source of the $\delta(x)$ contribution  in a twist-3 distribution. 
 
 As another example, twist-3 PDF, $e(k^+)$ and quasi-PDF, $e^{quasi}(k^z)$ are shown in FIG. \ref{fig:ek}. As the nucleon is boosted to the IMF $e^{quasi}(k^z)$ converges to $e(k^+)$. $e(k^+)$ contains a $\delta(k^+)$ term which corresponds to a non-scaling term in $e^{quasi}(k^z)$ at $k^z=0$ as $P^z\rightarrow \infty$. FIG.\ref{fig:ex} shows that, this component becomes a representation of $\delta(x)$ in $e^{quasi}(x)$ as the nucleon is boosted to the IMF.


\section{Violation Of Sum Rules}\label{section:sumrules}

In this section we investigate the sum rules involving twist-3 parton distributions and conclude that they are violated if the $\delta(x)$ contributions are not included. The point $x=0$ is not experimentally accessible in DIS  since it is not consistent with the Bjorken limit. However if  conclusions are drawn from the smooth behavior near $x=0$ about the behavior at $x=0$ the sum rules would appear to be violated. Nevertheless there  is no doubt in the validity of these sum rules since they are direct consequences of Lorentz invariance.  Therefore the violation of the sum rules from the experimental data would provide an indirect evidence on the existence of the Dirac delta functions.

Lorentz invariance of twist-3 GPDs imply that,
\begin{eqnarray}\label{t3Gsumrule}
\int_{-1}^1{}dx G_i (x,\xi,\Delta)=0, \hspace{1cm} \int_{-1}^1{}dx \widetilde{G}_i (x,\xi,\Delta)=0. 
\end{eqnarray}
 
If there is a $\delta(x)$  and if it is not  included, the Lorentz invariance of twist-3 GPDs would be violated as follows,
 
 \begin{eqnarray}\label{t3Gsumruleviolated} 
\lim _{\epsilon\rightarrow0}\int_{-1}^{\epsilon}{}dx G_i (x,\xi=0,\Delta)+\lim _{\epsilon\rightarrow0}\int^{1}_{\epsilon}{}dx G_i (x,\xi=0,\Delta)\neq 0, \\
\lim _{\epsilon\rightarrow0}\int_{-1}^{\epsilon}{}dx \widetilde{G}_i (x,\xi=0,\Delta)+\lim _{\epsilon\rightarrow0}\int^{1}_{\epsilon}{}dx \widetilde{G}_i (x,\xi=0,\Delta)\neq 0.
\end{eqnarray}
 
The Lorentz invariance relations in Eq.(\ref{t3Gsumrule}) can be regarded as non-forward generalization of Burkhardt-Cottingham sum rule involving the twist-3 PDF $g_T(x)$ \cite{Burkhardt:1970ti},
\begin{eqnarray}\label{BCsumrule}
\int_{-1}^1{}dx g_1 (x)=\int_{-1}^1{}dx g_T (x).
\end{eqnarray}

Since the left-hand side of Eq.(\ref{BCsumrule}) is the axial charge, the integral on the right-hand side is finite. If  the twist-3 PDF  $g_T (x)$ has a contribution proportional to $\delta(x)$ and $g_1(x)$ does not, experimental measurements would not be able to confirm the sum rule in Eq.(\ref{BCsumrule}) and claim its violation. The same argument applies to the sum rule in Eq.(\ref{hsumrule}), in which left-hand side is equal to the tensor charge.

\begin{eqnarray}\label{hsumrule}
\int_{-1}^1{}dx h_1 (x)=\int_{-1}^1{}dx h_L (x).
\end{eqnarray}
 
Another sum rule including a twist-3 PDF is the $\sigma$-term sum rule given by Eq.(\ref{esumrule}),  which provides a relation between quark mass $m$ and nucleon mass $M$,

\begin{eqnarray}\label{esumrule}
\int_{-1}^1{}dx e (x)=\dfrac{1}{2M}\langle P| \overline{\psi}(0)\psi(0)|P\rangle=\dfrac{d}{dm}M.
\end{eqnarray}

If $e(x)$ contains a $\delta(x)$, similary the sum rule in Eq.(\ref{esumrule}) is violated  if the $\delta(x)$ is not  included,

 \begin{eqnarray}\label{esumruleviolated} 
\lim _{\epsilon\rightarrow0}\int_{-1}^{\epsilon}{}dx e(x)+\lim _{\epsilon\rightarrow0}\int^{1}_{\epsilon}{}dx e(x)\neq\dfrac{d}{dm}M.
\end{eqnarray}

\section{Summary }\label{section:sumary}

We have investigated the twist-3 GPDs, $G_2$ and $\widetilde{G}_2$ using QTM and SDM. In both models these twist-3 GPDs exhibit discontinuities at the points $x=\pm\xi$. In the limit,  $\xi\rightarrow0$ the discontinuities of $G_2$ are divergent in both QTM and SDM. Therefore its ERBL region resembles a representation of a $\delta(x)$. However  the discontinuities of $\widetilde{G}_2$ behave differently in the two different models. While they diverge in SDM, they stay finite in the QTM  as $\xi\rightarrow0$. 

 In the forward limit, the twist-3 GPD, $\widetilde{G}_2$ reduces to twist-3 PDF $g_2(x)$ and the discontinuities grow into a $\delta(x)$ in SDM. Calculation of the quasi-PDF $g_2^{quasi}(k^z)$ reveals that  the $\delta(x)$ term corresponds to a component  that does not scale as the nucleon is boosted to the IMF.
 
  The $\delta(x)$ contribution is not unique to the case of $g_2(x)$ and all the other twist-3 PDFs contain a $\delta(x)$ in both QTM and SDM  only with the exception of $g_2 (x)$ in the QTM. These $\delta(x)$ terms are not related to the twist-2 (WW) parts of the twist-3 PDFs since model calculations show that none of the twist-2 PDFs contain such a term. Violations of the sum rules containing twist-3 PDFs and GPDs experimentally would provide an indirect evidence on the existence of these $\delta(x)$ contributions.

In generalized tadpole diagrams, i.e. diagrams where a subdiagram is connected to the rest of the diagram at a single vertex, the appearance of $\delta(x)$ terms is trivial. Since there is no momentum flowing through that subdiagram it is clear that the contribution to PDFs from the subdiagram does not scale in the IMF. However for the rainbow-like diagrams considered here it is not trivial but we demonstrated that for higher twist PDFs there still appears such a component which does not scale.

\section{ Appendix }\label{section:Appendix}

\subsection{$G_2$ In QTM}

The divergent part of $G_2$ is calculated as,

\begin{equation}\label{eq:G2int}
-ig^2 \int{}\dfrac{d^2k_{\perp}dk^-}{(2\pi)^4}\dfrac{k^-8(p^+)^2(1+x)}{\big[(k+\dfrac{\Delta}{2})^2-m^2+i\epsilon\big]\big[(k-\dfrac{\Delta}{2})^2-m^2+i\epsilon\big]\big[(p-k)^2-\lambda^2+i\epsilon\big]}.
\end{equation}

$k^-$ in the numerator of Eq.(\ref{eq:G2int}) can be replaced by the following expression,
\begin{equation}\label{G2kminus}
k^-=\dfrac{M^2}{2p^+}-\dfrac{[(p-k)^2-\lambda^2]}{2(p^+-k^+)}-\dfrac{(k^2_{\perp}+\lambda^2)}{2(p^+-k^+)}.
\end{equation}

The second term in Eq.(\ref{G2kminus}) cancels the propagator in the denominator in Eq.(\ref{eq:G2int}) leading to the following contribution which is nonzero only in the ERBL region, $-\xi<x<\xi$.

\begin{equation}\label{G2int}
ig^24p^+\dfrac{(1+x)}{(1-x)} \int{}\dfrac{d^2k_{\perp}dk^-}{(2\pi)^4}\dfrac{1}{\big[(k+\dfrac{\Delta}{2})^2-m^2+i\epsilon\big]\big[(k-\dfrac{\Delta}{2})^2-m^2+i\epsilon\big]}.
\end{equation}

The result of this integral is dominated by the following term which diverges as $\xi\rightarrow0$ yielding a representation of $\delta(x)$,

\begin{equation}\label{eq:G2div}
 -\dfrac{g^2}{(2\pi)^2 }\dfrac{(1+x)}{\xi(1-x)}\ln \Lambda_{\perp}.
\end{equation}

\subsection{$\widetilde{G}_2$ In QTM}

The divergent part of $\widetilde{G}_2$ is calculated as

\begin{equation}\label{eq:G2Tint}
-ig^2 \int{}\dfrac{d^2k_{\perp}dk^-}{(2\pi)^4}\dfrac{k^-8(p^+)^2 (x+\xi^2)}{\big[(k+\dfrac{\Delta}{2})^2-m^2+i\epsilon\big]\big[(k-\dfrac{\Delta}{2})^2-m^2+i\epsilon\big]\big[(p-k)^2-\lambda^2+i\epsilon\big]}.
\end{equation}

$k^-$ in the numerator of Eq.(\ref{eq:G2Tint}) can be replaced by the expression given by the Eq.(\ref{G2kminus}),
where the second term cancels the propagator in the denominator  leading to the contribution,

\begin{equation}
ig^24p^+\dfrac{(x+\xi^2)}{(1-x)} \int{}\dfrac{d^2k_{\perp}dk^-}{(2\pi)^4}\dfrac{1}{\big[(k+\dfrac{\Delta}{2})^2-m^2+i\epsilon\big]\big[(k-\dfrac{\Delta}{2})^2-m^2+i\epsilon\big]}
\end{equation}

The result of this integral is dominated by the following term,

\begin{equation}
-\dfrac{g^2}{(2 \pi)^2 }\dfrac{(x+\xi^2)}{\xi(1-x)}\ln \Lambda_{\perp}.
\end{equation}

Due to the apprearence of $\xi^2$ in the numerator this expression is finite as $\xi\rightarrow0$ unlike the expression in Eq.(\ref{eq:G2div})

\subsection{$g_T(x)$ In SDM}

The parametrization involving $g_T(x)$ is given by,
\begin{eqnarray}\label{gTapp}
\int{}\dfrac{d\lambda}{2\pi}e^{i \lambda x}\langle P,S|\overline{q}(0)\gamma^{\mu}\gamma_5q(\lambda n)|P,S\rangle=2\big\{g_1(x) \hat{p}^{\mu}(S \cdot \hat{n})+g_T(x)S_{\bot}^{\mu}+ M^2 g_3(x) \hat{n}^{\mu}(S \cdot \hat{n}) \big\}.
\end{eqnarray}
 
$g_T(x)$ is extracted by using $\gamma^{\mu}=\gamma^{\perp}$ in Eq.(\ref{gTapp}),

\begin{eqnarray}\label{gTp}
\int{}\dfrac{d\lambda}{2\pi}e^{i \lambda x}\langle P,S|\overline{q}(0)\gamma^{\perp}\gamma_5q(\lambda n)|P,S\rangle=2g_T(x)S^{\perp}.
\end{eqnarray}

The left-hand side of Eq.(\ref{gTp}) is written using SDM,
\begin{eqnarray}\label{gTint}
g_T(x)S^{\perp}=\dfrac{ig^2}{2}\int{}\dfrac{d^4k}{(2 \pi)^4}\delta(k^+-xP^+) \overline{u}(p) \dfrac{(k\!\!\!/+m)}{(k^2-m^2+i\epsilon)} \gamma^{\perp}{\gamma_5}\dfrac{(k\!\!\!/+m)}{(k^2-m^2+i\epsilon)}u(p)\dfrac{1}{\big[(p-k)^2-\lambda^2+i\epsilon\big]}.
\end{eqnarray}

 Using the following identities,
\begin{eqnarray}\label{G1}
 \overline{u}(P) u(P)=2M,
 \end{eqnarray}
\begin{eqnarray}\label{G2}
 \overline{u}(P) \gamma^{\mu}u(P)=2P^{\mu},
 \end{eqnarray}
\begin{eqnarray}\label{G3}
\overline{u}(P) \gamma_5u(P)=0,
 \end{eqnarray}
\begin {eqnarray}\label{G4}
  \overline{u}(P) \gamma^{\mu}\gamma_5u(P)=2S^{\mu},
 \end{eqnarray}
\begin{eqnarray}\label{G5}
  \overline{u}(P) \gamma^{+}\gamma^{\perp}\gamma_5u(P)=\dfrac{2P^+}{M}  S^{\perp},
 \end{eqnarray}
 \begin{eqnarray}\label{G6}
  \overline{u}(P) \gamma^{-}\gamma^{\perp}\gamma_5u(P)=\dfrac{2P^-}{M} S^{\perp}=\dfrac{M}{P^+} S^{\perp},
 \end{eqnarray}
 \begin{eqnarray}\label{G7}
  \overline{u}(P) \gamma^{-}\gamma^{+}\gamma_5u(P)=\dfrac{2M}{P^+} S^{+},
 \end{eqnarray}
 the numerator of the integrand in Eq.(\ref{gTint}) is calculated,

\begin {eqnarray}\label{gTnum}
\overline{u}(P)(k\!\!\!/+m)\gamma^{\perp}\gamma_5(k\!\!\!/+m)u(P)=\Big[4k^-P^+(x+\dfrac{m}{M})+2k_{\perp}^2+2mM(x+\dfrac{m}{M})\Big]S^{\perp}.
\end{eqnarray}

As in the case for $G_2$ and $\widetilde{G}_2$, when $k^-$ in Eq.(\ref{gTnum}) is replaced by the following expression,
\begin{eqnarray}\label{kminus}
k^-=\dfrac{M^2}{2P^+}-\dfrac{[(P-k)^2-\lambda^2]}{2(P^+-k^+)}-\dfrac{(k^2_{\perp}+\lambda^2)}{2(P^+-k^+)},
\end{eqnarray}

 the second term  cancels the propagator in the denominator of Eq.(\ref{gTint})  and leads to the contibution,

\begin{eqnarray}\label{gTintegral}
-\dfrac{ig^2}{(2\pi)^2}\dfrac{(x+\dfrac{m}{M})}{(1-x)}\int{}\dfrac{d^2k_{\perp}}{(2\pi)^2} \dfrac{dk^-}{(k^2-m^2+i\epsilon)^2}=
\dfrac{g^2}{16\pi^2 }\dfrac{(x+\dfrac{m}{M})}{(1-x)} \ln(\dfrac{\Lambda^2_{\perp}+m^2}{m^2})\delta(x),
\end{eqnarray}

where  $\Lambda_{\perp}$  is the transverse momentum cut off and,

\begin{eqnarray}\label{w}
\omega=-x(1-x)M^2+(1-x)m^2+x\lambda^2.
\end{eqnarray}

\subsection{$g_1(x)$ In SDM }

$g_1(x)$ is extracted by using $\gamma^{+}=\gamma^{\perp}$ in Eq.(\ref{gTapp}),

\begin{eqnarray}\label{g1}
\int{}\dfrac{d\lambda}{2\pi}e^{i \lambda x}\langle P,S|\overline{q}(0)\gamma^{+}\gamma_5q(\lambda n)|P,S\rangle&=2g_1(x) S^+
\end{eqnarray}

The left-hand side of Eq.(\ref{g1}) is written using SDM,

\begin{eqnarray}\label{g1int}
g_1(x)S^+=\dfrac{ig^2}{2}\int{}\dfrac{d^4k}{(2 \pi)^4}\delta(k^+-xP^+) \overline{u}(p) \dfrac{(k\!\!\!/+m)}{(k^2-m^2+i\epsilon)} \gamma^+{\gamma_5}\dfrac{(k\!\!\!/+m)}{(k^2-m^2+i\epsilon)}u(p)\dfrac{1}{\big[(p-k)^2-\lambda^2+i\epsilon\big]}.
\end{eqnarray}

 The numerator of the integrand in Eq.(\ref{g1int}) is calculated using the identities given by Eqs.(\ref{G1})-(\ref{G7}),

\begin{eqnarray}\label{g1num}
\overline{u}(p)(k\!\!\!/+m)\gamma^+\gamma_5(k\!\!\!/+m)u(p)&=-4k^{+2}S^--[2k_{\perp}^2-2m^2-4mMx]S^+\\ \nonumber
&=2[(m+xM)^2-k_{\perp}^2]S^+.
\end{eqnarray}

Using Eq.(\ref{g1num})  in Eq.(\ref{g1int}) leads to,

\begin{eqnarray}
g_1(x)=&\dfrac{g^2}{16\pi^2}(1-x)\bigg\{\dfrac{[\omega+(m+xM)^2]}{k_{\perp}^2+\omega}+\ln(k_{\perp}^2+\omega)\bigg\}_{k_{\perp}^2=0}^{\Lambda_{\perp}^2}
\end{eqnarray}

where  $\omega$ is given by the Eq.(\ref{w}). Similar to other twist-2 PDFs, the numerator in Eq.(\ref{g1num}) does not have a term involving $k^-$ and therefore the cancellation of the propagator which leads the $\delta(x)$ contribution does not occur.

{\bf Acknowledgements:}
We thank C. Lorc\'e, A. Metz and B. Pasquini for helpful discussions.
This work was partially supported by the DOE under grant number 
DE-FG03-95ER40965. 



\end{document}